\documentclass[11pt,letter]{article}
\usepackage[utf8]{inputenc}
\usepackage{special,times,graphicx}
\usepackage{authblk}
\usepackage{blindtext}
\usepackage{hyperref}
\usepackage[square,numbers]{natbib}
\usepackage[english]{babel}
\usepackage{float} 
\usepackage{url}

\title{Mapping county-level mobility pattern changes in the United States in response to COVID-19 }
\author{Song Gao\thanks{Corresponding Email: song.gao@wisc.edu} $^1$, Jinmeng Rao$^1$, Yuhao Kang$^1$, Yunlei Liang$^1$, Jake Kruse$^1$\\$^1$GeoDS Lab, Department of Geography, University of Wisconsin-Madison, WI 53706, USA}

\begin{document}
\maketitle

\begin{abstract}
To contain the COVID-19 epidemic, one of the non-pharmacological epidemic control measures is reducing the transmission rate of SARS-COV-2 in the population through social distancing. An interactive web-based mapping platform that provides timely quantitative information on how people in different counties and states reacted to the social distancing guidelines was developed by the GeoDS Lab @UW-Madison with the support of the National Science Foundation RAPID program. The web portal integrates geographic information systems (GIS) and daily updated human mobility statistical patterns (median travel distance and stay-at-home dwell time) derived from large-scale anonymized and aggregated smartphone location big data at the county-level in  United States. It aims to increase risk awareness of the public, support data-driven public health and governmental decision-making, and help enhance community responses to the COVID-19 pandemic.
\end{abstract}  

\section{Introduction}
\begin{figure}[h]
	\centering
	\includegraphics[width=0.95\textwidth]{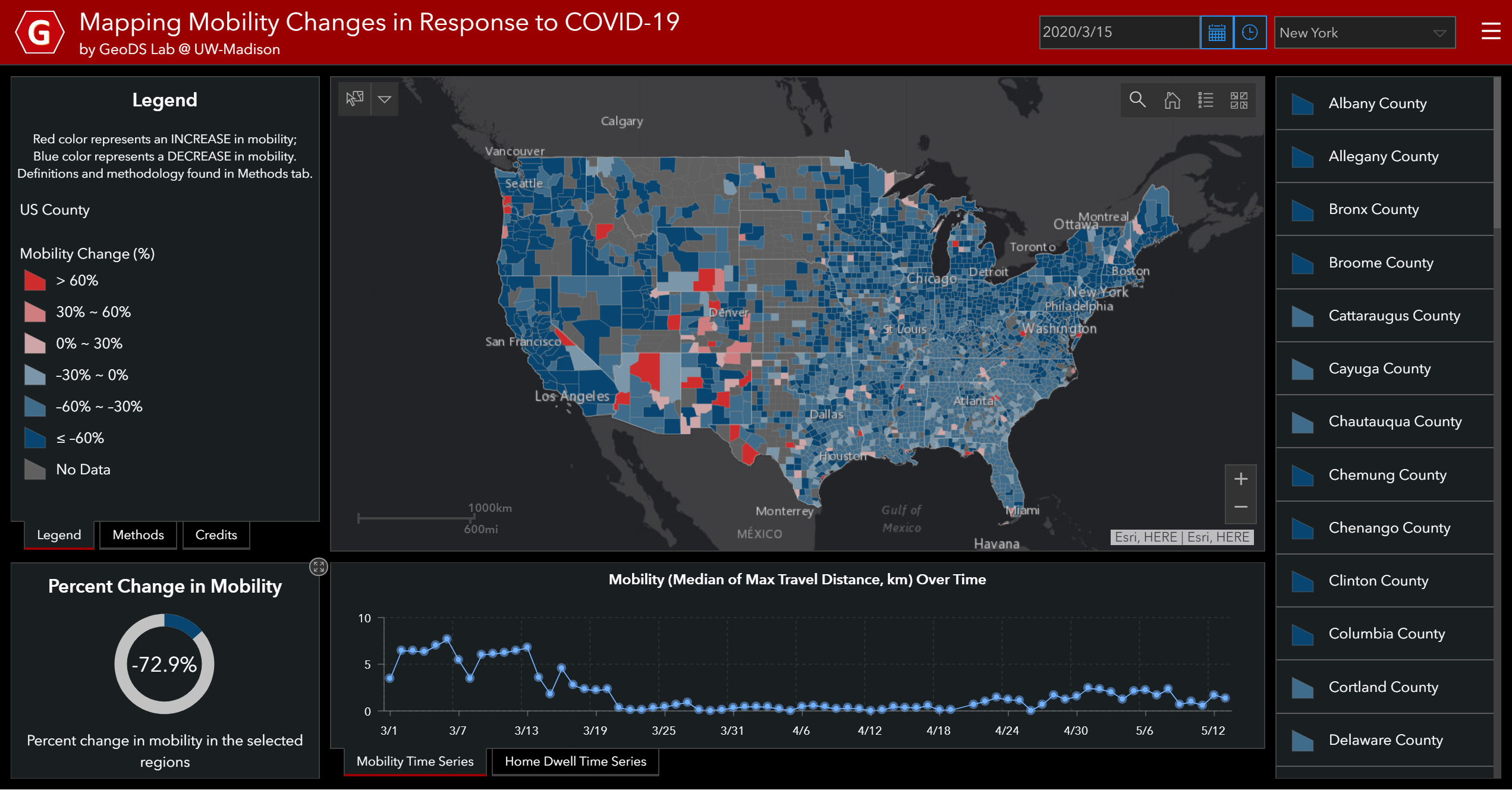}
	\caption{The web interface of the mapping platform.}
	\label{fig:geods_webportal}
\end{figure}

The coronavirus disease (COVID-19) pandemic is a global threat with escalating health,
economic and social challenges. As of May 12, 2020, there had been 1,342,594 total
confirmed cases and 80,820 total deaths in the U.S. according to the reports of the U.S. Centers for
Disease Control and Prevention (CDC) \footnote{\url{https://www.cdc.gov/coronavirus/2019-ncov/cases-updates/cases-in-us.html}}. To contain the COVID-19 epidemic, one of the non-pharmacological epidemic control measures is reducing the transmission rate of SARS-COV-2 in the population through (physical) social distancing. An interactive web-based mapping platform\footnote{\url{https://geods.geography.wisc.edu/covid19/physical-distancing/}} (as shown in Fig. \ref{fig:geods_webportal}) that provides timely quantitative information on how people in different counties and states reacted to the social distancing guidelines and followed stay-at-home mandates was developed by the GeoDS Lab at the University of Wisconsin-Madison with the support of the National Science Foundation. It integrates geographic information systems (GIS) and daily updated human mobility statistical patterns (median travel distance and stay-at-home dwell time) derived from large-scale anonymized and aggregated smartphone location big data at the county-level in United States \cite{zhou2020covid,descarteslabs2020mobility,prestby2019understanding,liang2020calibrating}. This mobility dashboard aims to increase risk awareness of the public, support data-driven decision-making, and help enhance community responses to the COVID-19 pandemic. 

With the rapid development of information, communication, and technologies, new data acquisition and assessment methods are available to help evaluate the risk awareness of epidemic transmission and geographic spreading from the community perspective. The online mobility mapping platform can help effectively monitor not only the spatiotemporal changes of travel distance and stay-at-home time of people as well as the effect of social distancing policies on human movement behaviors. It can also inform the analysis of potential influential factors on the compliance with COVID-19 social distancing orders such as self-report beliefs, social disparities, political and geographic contexts in risk communication and public-health interventions \cite{gollwitzer2020connecting,allcott2020polarization,painter2020political,kavanagh2020association}. Recent studies have utilized various sources (e.g., Google mobility reports) of mobile location big data to quantify mobility change patterns and its relation to COVID-19 spread, air pollution, and economic impacts \cite{andersen2020early, dasgupta2020quantifying,zhang2020interactive,ghader2020observed,liang2020urban,gao2020mobile}. In the following, we first introduce the data sources and the design principles and techniques to develop the mobility tracking Web portal in detail. Then, we describe the observed human mobility changes in the U.S. in three periods: before stay-at-home orders, after stay-at-home mandates in place, and the economy partial reopening period.

\section{Methods}
\subsection{Data Sources}
In response to the rapid spread of the SARS-COV-2 virus, as of April 7, 2020, 42 states and Washington D.C. in the U.S. have issued the statewide ``stay-at-home" or ``safer-at-home'' orders\footnote{\url{https://www.nytimes.com/interactive/2020/us/coronavirus-stay-at-home-order.html}}. People generally tend to change their travel behavior to slow down the spread of the novel coronavirus but geographic and temporal heterogeneity still exists. To quantitatively understand how people reacted to the stay-at-home orders imposed during the COVID-19 outbreak, human mobility changes are considered in terms of changes in median travel distance and stay-at-home dwell time.

We first use the U.S. mobility data released by the Descartes Labs \cite{descarteslabs2020mobility} to map the daily travel changes. The individual mobility is measured by the maximum travel distance (km) to a location from the initial location of the day given a unique mobile device (i.e., individual max-distance mobility). To investigate the mobility changes in a geographic region, a baseline was first determined \cite{descarteslabs2020mobility}, which was defined as the median of the max-distance mobility on the weekdays between 2/17/2020 and 3/7/2020 in the specified region (i.e., by county). By comparing daily mobility with baseline mobility, we can measure how people in each county react to COVID-19 by reducing their daily travel distance:

\textbf{Median Travel Distance}: The median of the individual daily maximum travel distance for all location samples in the specified region.

\textbf{Percent Change in Mobility}: The percentage change in the daily median mobility from the baseline.

\begin{figure}[h]
	\centering
	\includegraphics[width=0.95\textwidth]{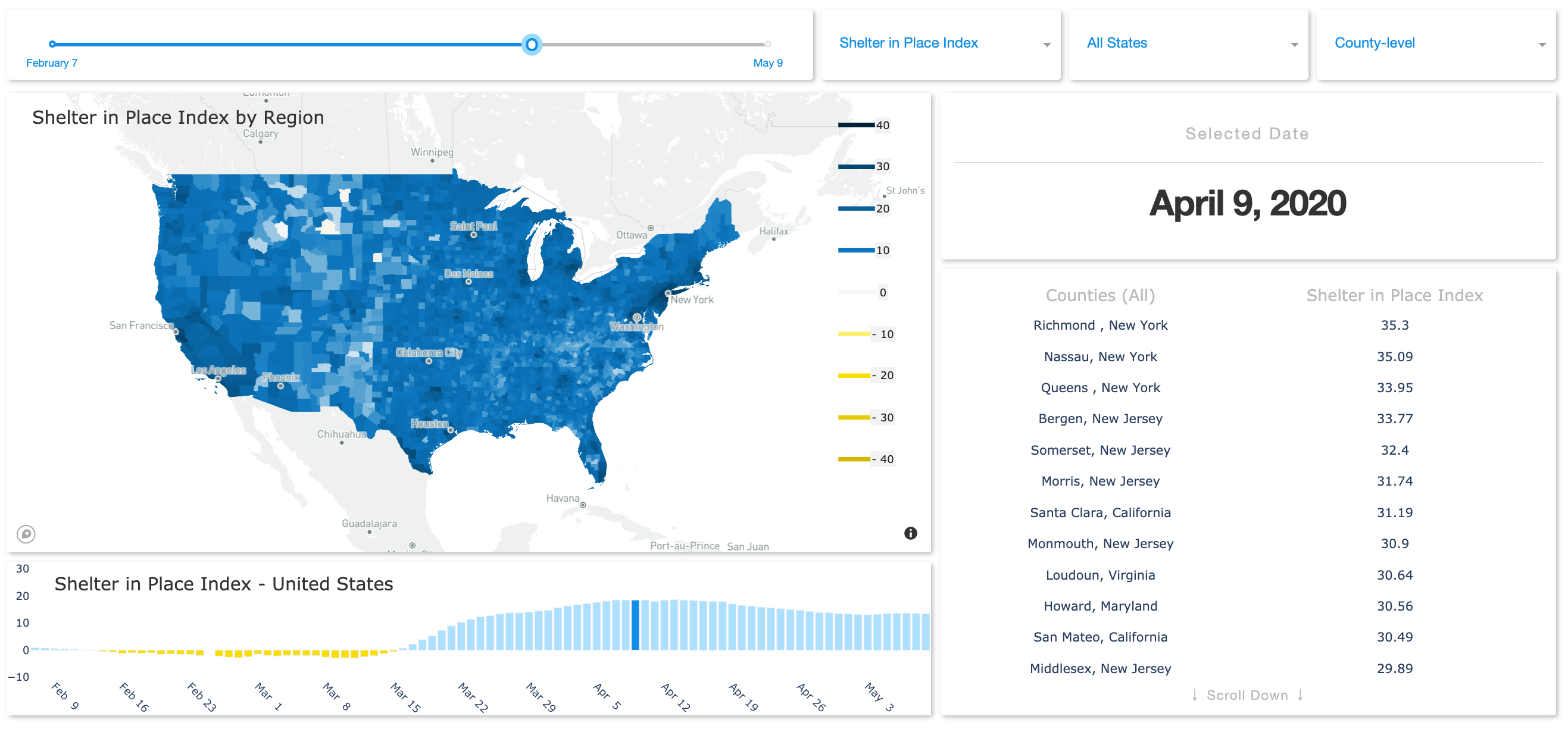}
	\caption{The stay-at-home dwell time data provided by SafeGraph.}
	\label{fig:safegraph_shelter_in_place}
\end{figure}

In addition, we use the SafeGraph ``Social Distancing Metrics'' data\footnote{\url{https://docs.safegraph.com/docs/social-distancing-metrics}} to compute the \textbf{Median of Stay-at-Home Dwell Time} for understanding people's compliance to the stay-at-home orders in each county across the nation. Fig. \ref{fig:safegraph_shelter_in_place} shows the percent of the population staying home (sheltering in place) given a particular date. Data preprocessing is needed to determine the common nighttime location (i.e., home location) of each mobile device over a 6-week period to a Geohash-7 granularity (153m x 153m). The devices are aggregated by home census block group and the metrics are further computed to the county level. SafeGraph data provide unique and valuable insights into the foot-traffic changes to large-scale businesses and consumer POIs across the nation \cite{safegraph}. 
To measure stay-at-home dwell time, the home place for each individual is identified and the hours/minutes for all sampled devices staying at that home place across the day are summed up. To measure dynamic changes in home dwell time, the median home dwell time for all observed devices is computed in different geographical units (e.g., census block groups, counties, and states).

\subsection{System Design}

\begin{figure}[h]
	\centering
	\includegraphics[width=0.95\textwidth]{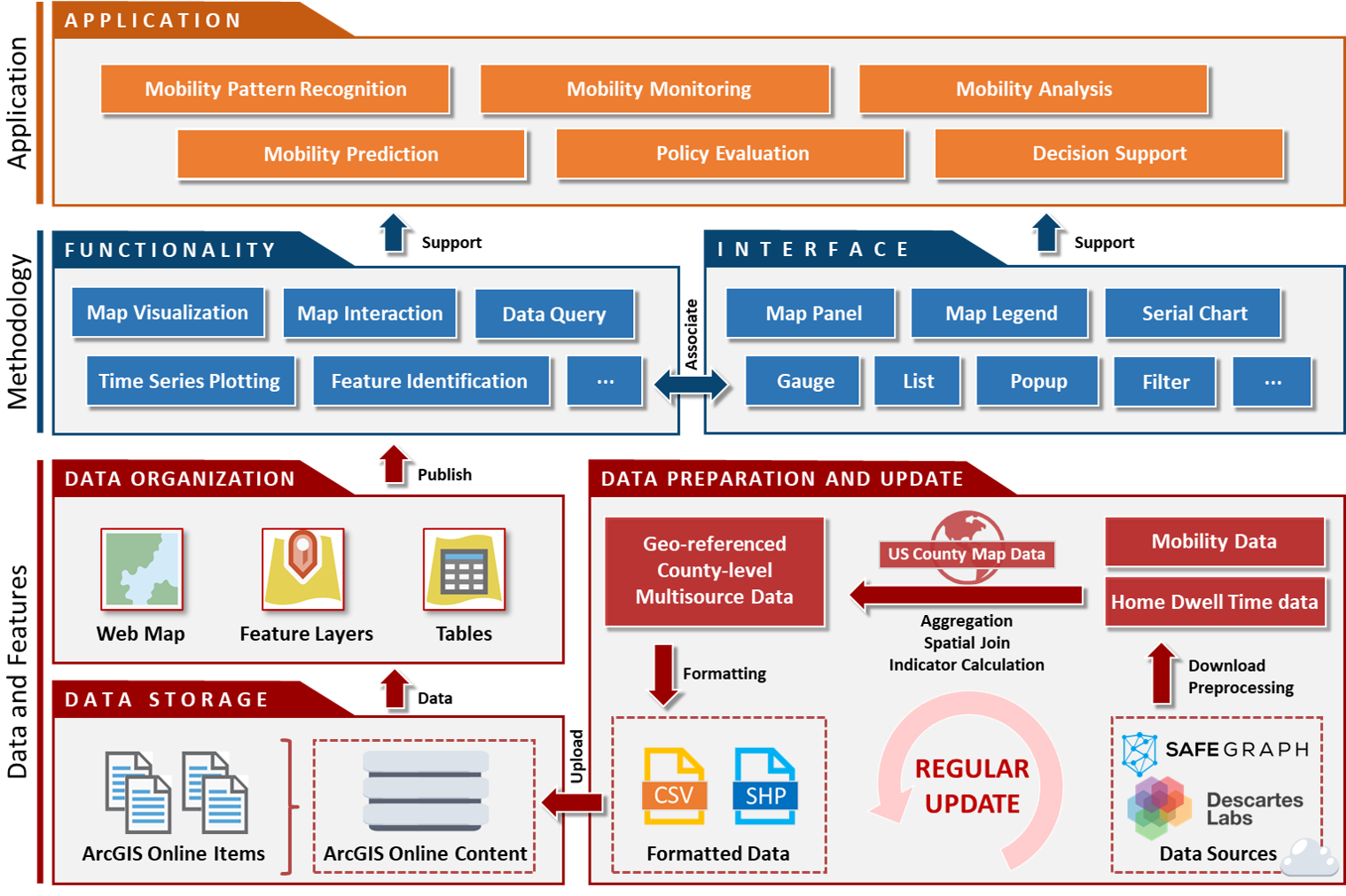}
	\caption{The mobility tracking system design diagram.}
	\label{fig:system_design}
\end{figure}
 
The interactive web mapping platform is designed and developed using the ArcGIS Operational Dashboards\footnote{\url{https://doc.arcgis.com/en/dashboards/}}. It integrates data streams with maps, time series plots, gauges, and other visual elements to represent the dynamic status of human mobility patterns comprehensively. The system design diagram for the mobility tracking dashboard is shown in Fig. \ref{fig:system_design}. 

Our mobility tracking dashboard consists of three layers: \textit{Data and Feature layer, Methodology layer, and Application layer}. At the Data and Feature layer, we first acquire the abovementioned travel distance and home dwell time mobility data from the two data providers and preprocess them (e.g., data cleaning, data quality check). Then we aggregate the processed datasets to the county level and match them to the U.S. county map (using the spatial join operator). Mobility indicators such as the median travel distance mobility index and the median home dwell time are calculated during this process. Finally, these datasets are formatted and uploaded to ArcGIS Online Content as web items, which are then organized as web maps, feature layers, and time series tables for publication. The whole process is iterating regularly to keep the data up to date.

At the Methodology layer, we apply the functionality design (e.g., map visualization, data query, time-series data plotting) and the interface design (e.g., component design, interaction design). These two parts are internally associated to help users better access, understand, and interact with the data. At the Application layer, the system can further support various mobility-related applications such as mobility pattern recognition, monitoring, analysis, modeling, prediction, and decision support. For instance, the mobility change data product has been used in epidemic modeling \cite{chen2020mitigating,gao2020mobile} and in air pollution association analysis \cite{liang2020urban}. Furthermore, to optimize the mobile user interface and improve user experience \cite{rothcv}, we employed another design tool -- the ArcGIS Experience Builder\footnote{\url{https://www.esri.com/en-us/arcgis/products/arcgis-experience-builder/overview}} to construct mapcentric apps and display them on a scrolling and multi-panel screen for mobile devices.

\section{Mobility Changes and Insights}
On the mobility tracking map, the red color counties represent an INCREASE in daily mobility compared with the abovementioned baseline statistic, while blue color counties represent a DECREASE in mobility. The color saturation reflects the degree of changes. The darker the more mobility changes in a county. In the following, we discuss the observed mobility patterns in three periods.

\par \noindent \textit{\textbf {Before stay-at-home orders}}
\par
\begin{figure}[H]
	\centering
	\includegraphics[width=0.95\textwidth]{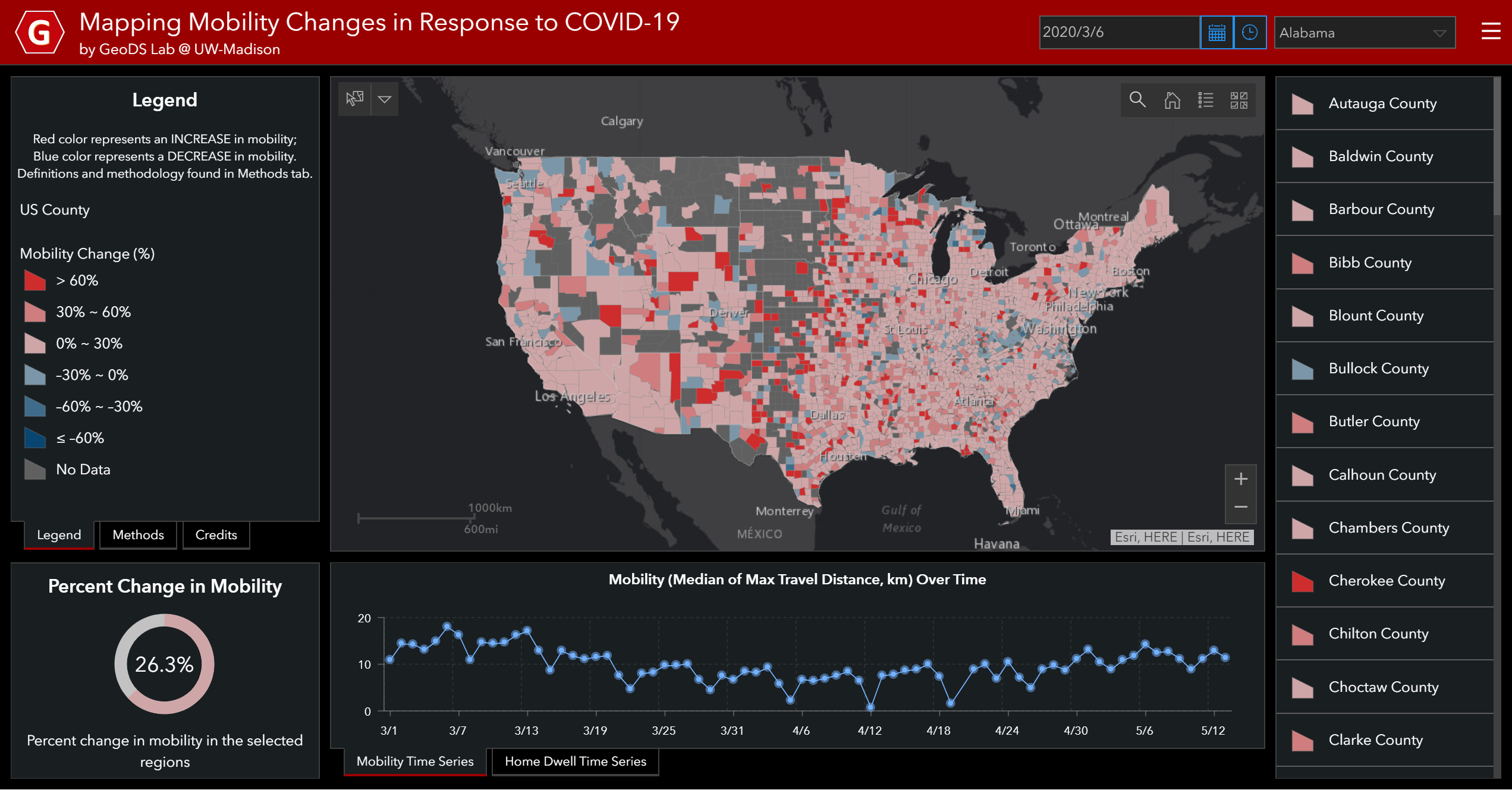}
	\caption{The mobility change patterns in the U.S. on March 6, 2020.}
	\label{fig:march6}
\end{figure}

 As shown in Fig. \ref{fig:march6}, the median travel distance in most counties (except some counties such as the King County, Washington, one of the known epidemic outbreak centers in the U.S. \cite{mcmichael2020covid}) across the nation largely increased on March 6, 2020 as expected on a normal spring travel date as usual if this pandemic didn't come out. 

\par \noindent \textit{\textbf {During stay-at-home mandates}}

\begin{figure}[H]
	\centering
	\includegraphics[width=0.95\textwidth]{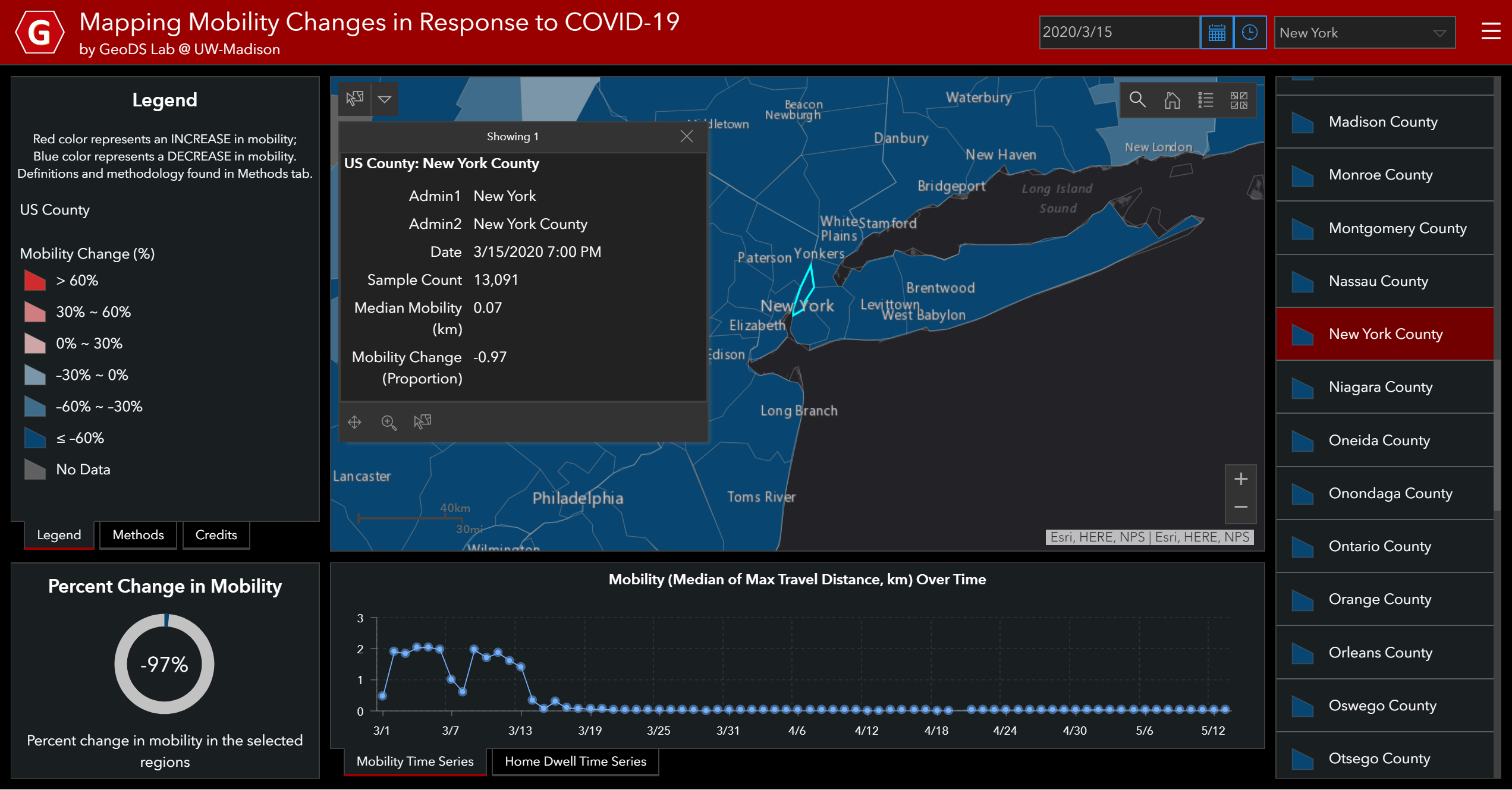}
	\caption{The mobility change patterns in New York on March 15, 2020.}
	\label{fig:NY}
\end{figure}

\begin{figure}[H]
	\centering
	\includegraphics[width=0.95\textwidth]{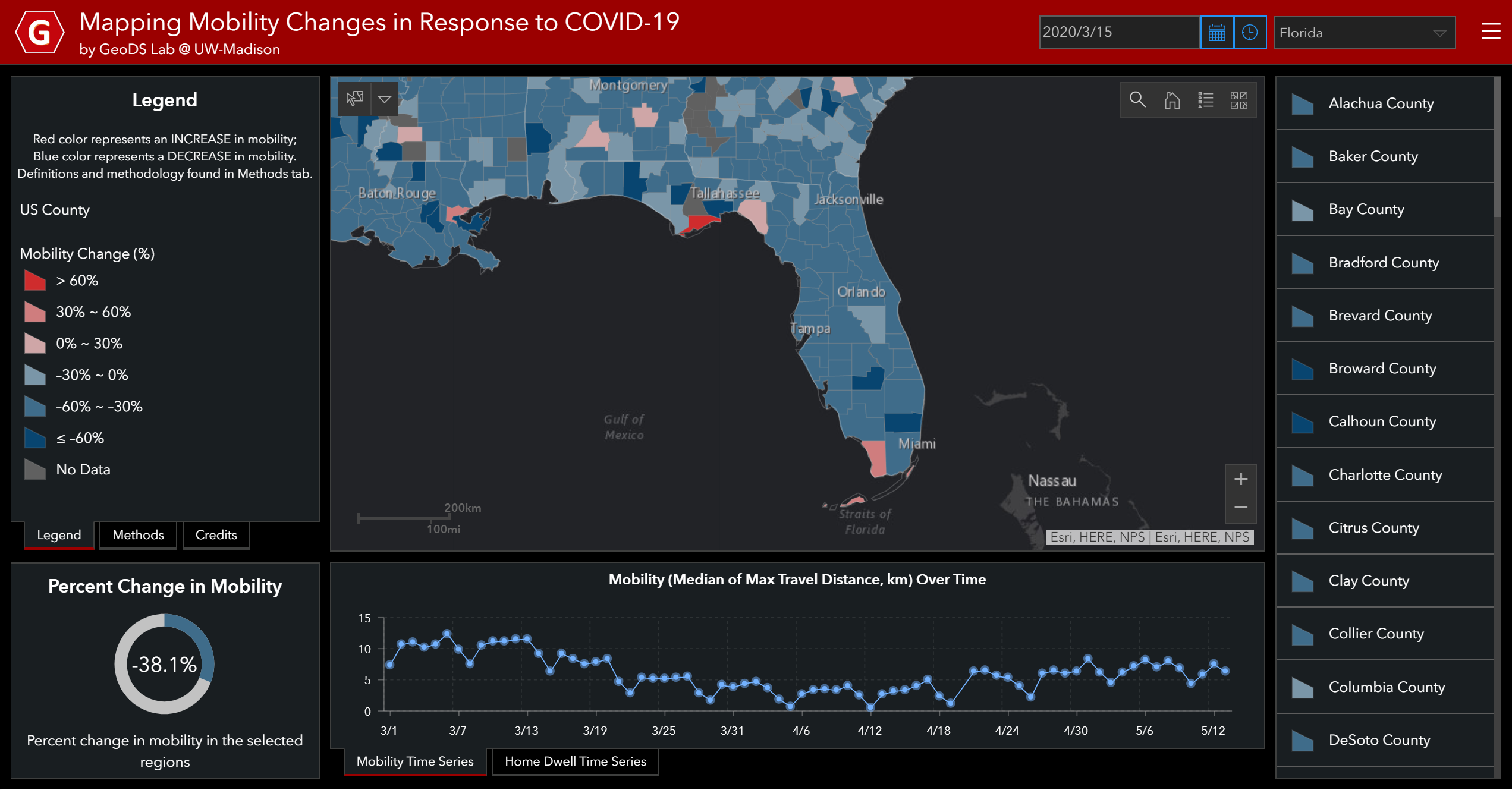}
	\caption{The mobility change patterns in Florida on March 15, 2020.}
	\label{fig:FL}
\end{figure}

\begin{figure}[H]
	\centering
	\includegraphics[width=0.95\textwidth]{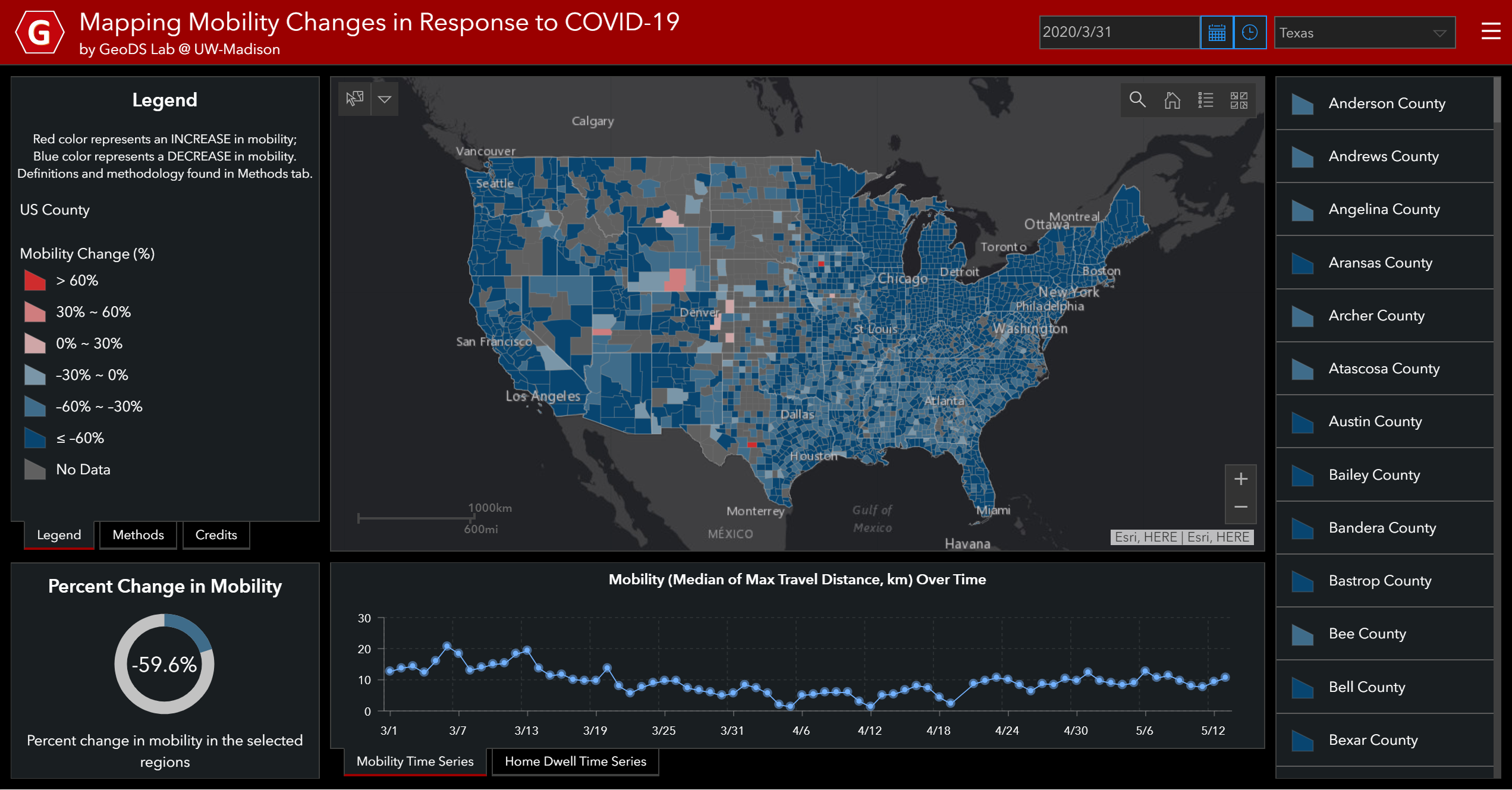}
	\caption{The U.S. mobility change patterns on March 31, 2020.}
	\label{fig:march31}
\end{figure}

\begin{figure}[H]
	\centering
	\includegraphics[width=0.95\textwidth]{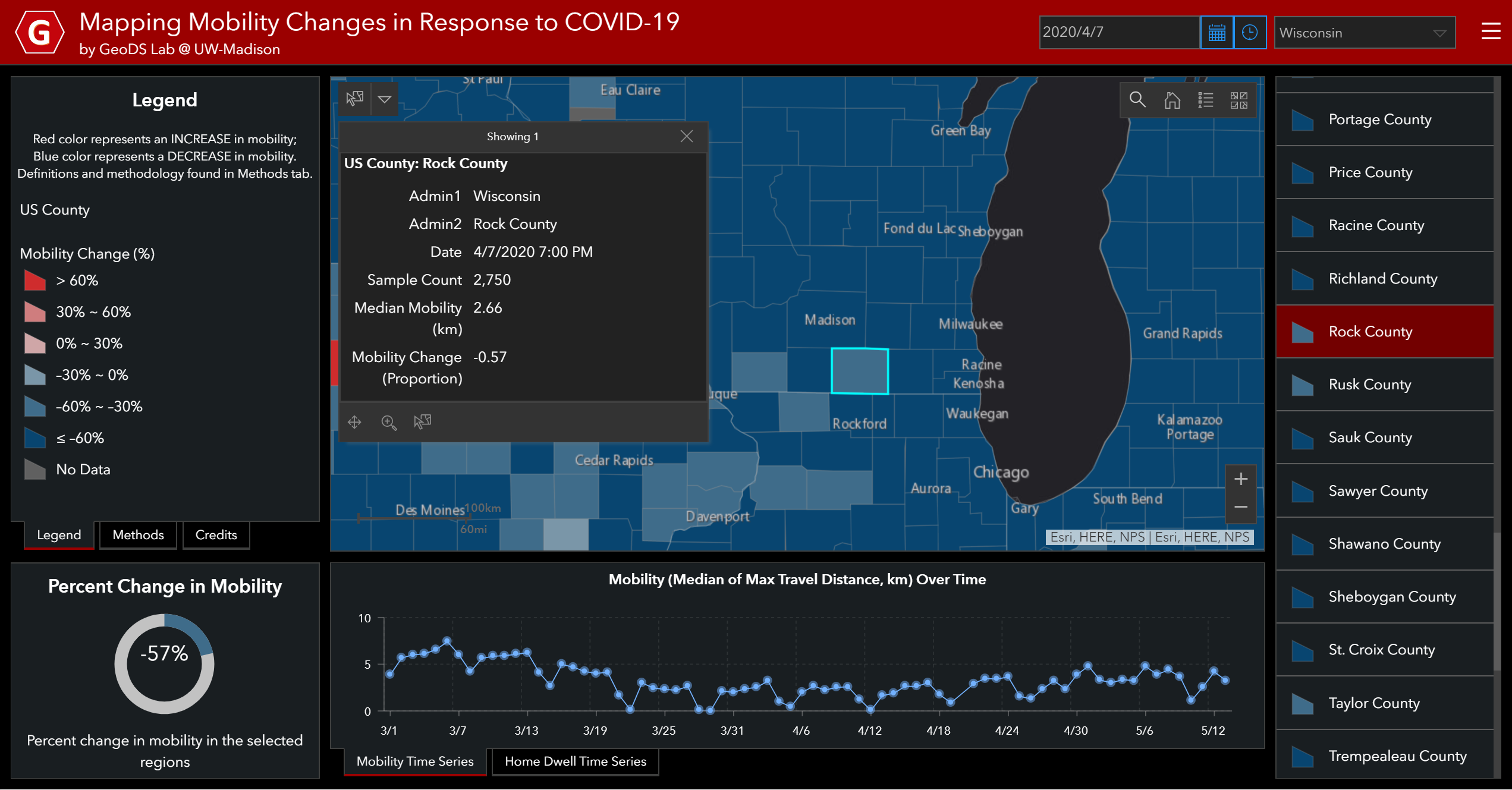}
	\caption{The mobility change patterns in Wisconsin on April 7, 2020.}
	\label{fig:WI}
\end{figure}

However, as of March 15, 2020, two days after the U.S. Federal Government announced the national emergency, people in most states in the Pacific Coast, Midwest, and East Coast had reacted actively to the social distancing guidelines and reduced their daily mobility (in Fig. \ref{fig:geods_webportal}). But the adherence to the stay-at-home orders is geographically heterogeneous across the nation. By zooming into the New York state (in Fig. \ref{fig:NY}), its median max-distance reduced to less than 0.1 km and decreased about 73\% of that compared to the baseline. Despite warnings from health experts that drastic control measures are needed to slow the spread of the virus, people in the Franklin County (Florida) and the Monroe County (Florida) as shown in Fig. \ref{fig:FL}, and many other counties in Arizona, New Mexico, Colorado, Utah and Wyoming still have increased daily mobility patterns (Fig. \ref{fig:geods_webportal}) since there was no statewide lock-down order yet. As of March 31, 2020, most states in the U.S. had issued fully shelter-in-place or partially safer-at-home and no large-group gatherings orders \cite{businessinsider}, asking residents to stay at home and go out only for essential services, such as grocery shopping and medical cares. Thus, the spatial distribution of the mobility changes was dominantly in blue (decreased mobility) on that day except for some counties (as shown in Fig. \ref{fig:march31}), including Sutton County (Texas), Carbon County (Wyoming), Big Horn County (Montana), Millard County (Utah), Yuma, Prowers, and Lincoln counties (Colorado), and Pocahontas and Decatur counties in Iowa. The county/state mobility patterns may change over time if there will be some inevitable gathering events (e.g., in-person voting in presidential primary elections and protests). For example, when casting a ballot in Wisconsin’s April 7 election, crowds lined up at polling places across the state — including thousands funneled to five sites in Milwaukee. A small increase of median travel distance from 1.1 kilometers on Monday to 1.9 km in the Waukesha County (within the metropolitan area of Milwaukee), and longer-distance travels in Lafayette County (from to 2.7 km April 6 to 4.5 km April 7) and in Rock County (2 km to 2.66 km) were observed on that day (in Fig. \ref{fig:WI}). Later, 71 people who went to the primary polls on April 7 were tested positive for the SARS-COV-2 virus\footnote{\url{http://madison.com/wsj/news/local/health-med-fit/71-people-who-went-to-the-polls-on-april-7-got-covid-19-tie-to/article_ef5ab183-8e29-579a-a52b-1de069c320c7.html}}. A study found a statistically significant association between in-person voting and the spread of COVID-19 cases two to three weeks after the election \cite{cotti2020relationship}. 

\par \noindent \textit{\textbf {Reopening period}}

\begin{figure}[H]
	\centering
	\includegraphics[width=0.95\textwidth]{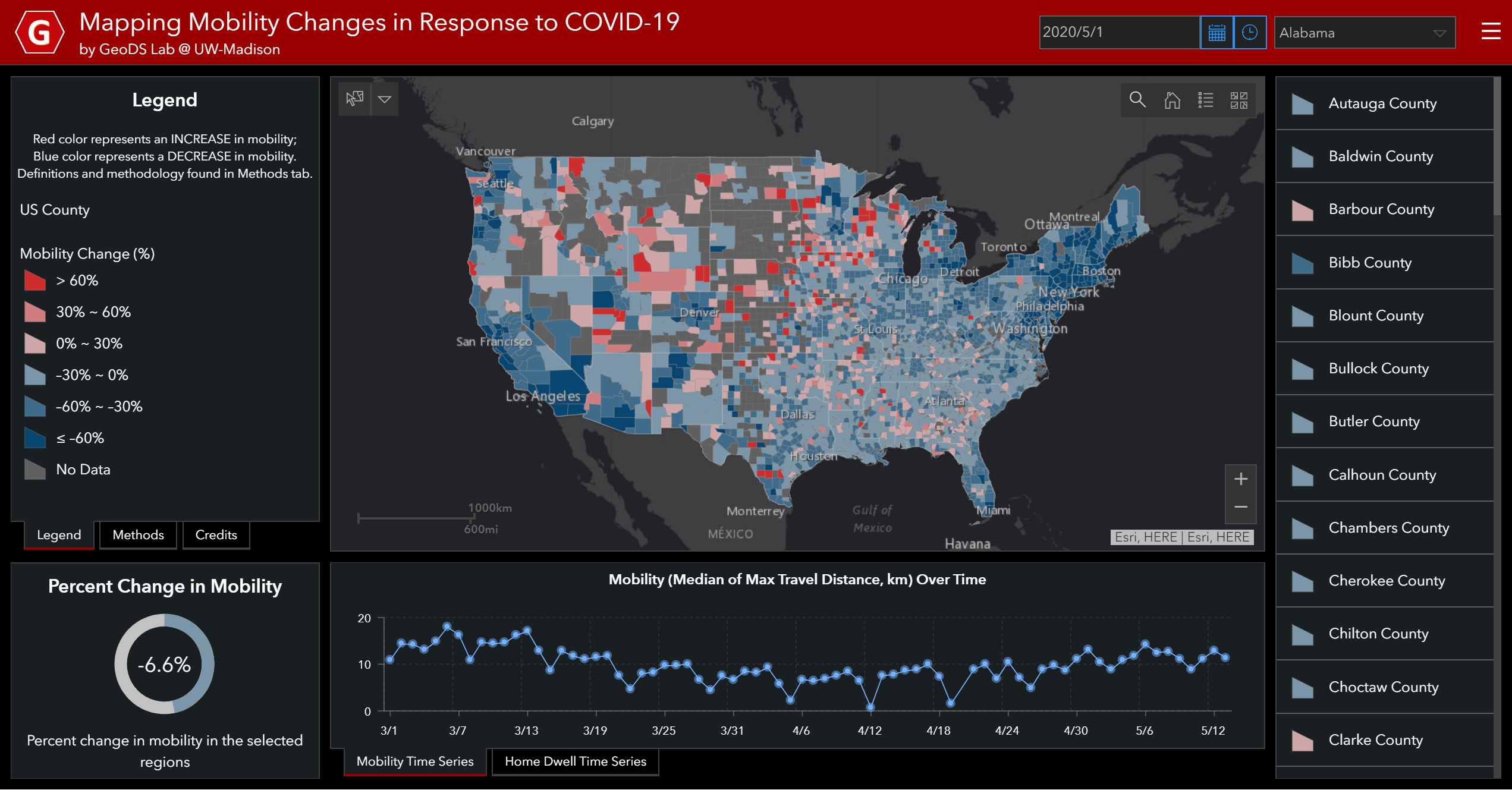}
	\caption{The U.S. mobility change patterns on May 1, 2020.}
	\label{fig:may1}
\end{figure}

Since early May, 2020, when several states began to lift their stay-at-home and non-essential business restrictions\footnote{\url{https://www.nytimes.com/interactive/2020/us/states-reopen-map-coronavirus.html}}, more and more people returned to public life and the median travel distance bounced up (in Fig. \ref{fig:may1}) after keeping at the minimum value for a while during the stay-at-home period. Such an upward mobility trend is expected to continue as more states partially reopen their economy. 

\begin{figure}[H]
	\centering
	\includegraphics[width=0.95\textwidth]{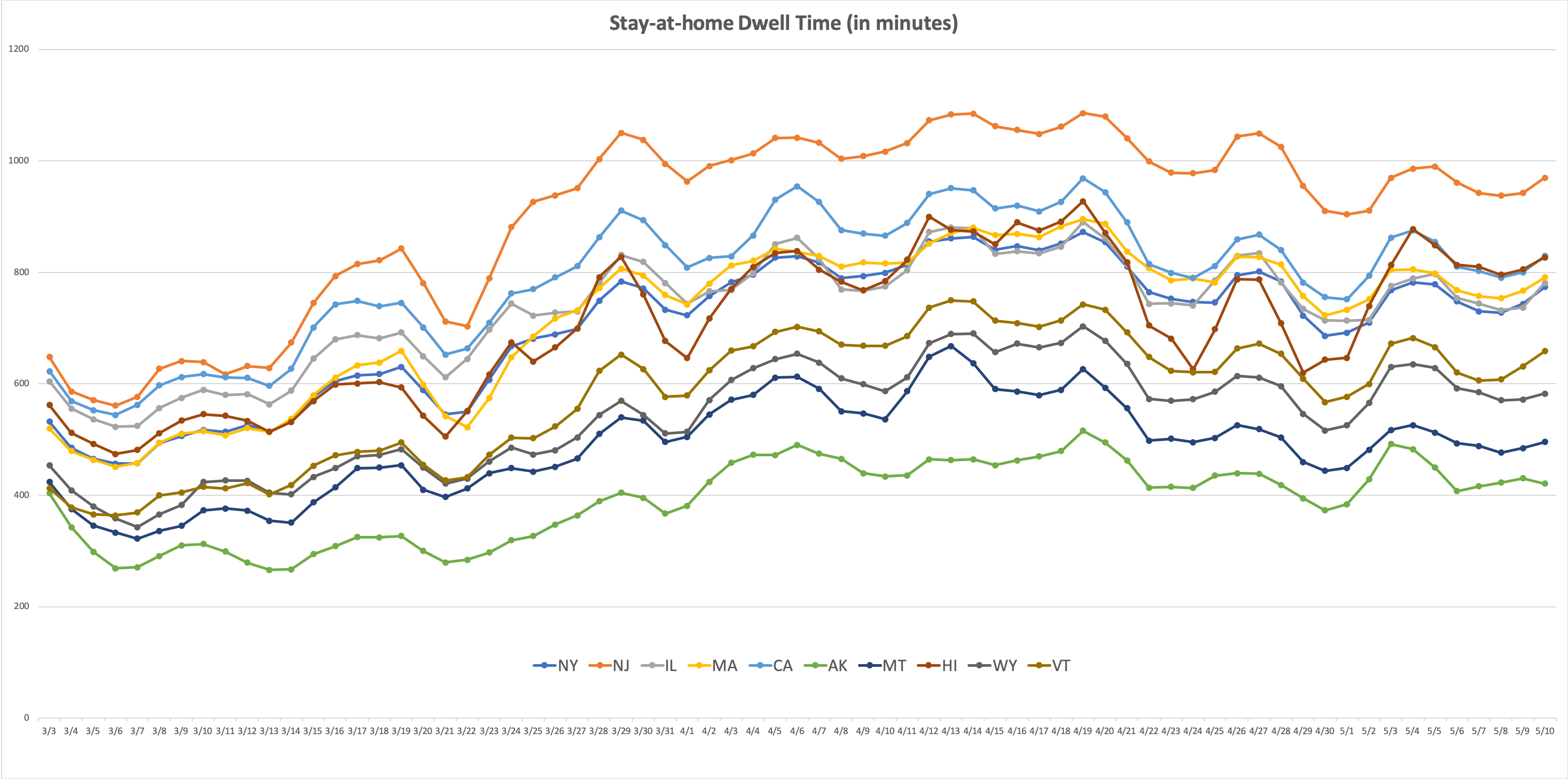}
	\caption{The three-day moving average of the median stay-at-home dwell time changes for the five most and the five least infected states (Data Source: SafeGraph).}
	\label{fig:stayathometime}
\end{figure}

In addition, by examining the state-specific stay-at-home dwell time temporal patterns (in Fig. \ref{fig:stayathometime}), we found that the three-day moving averages of the median stay-at-home dwell time generally are higher for people residing in the top 5 most infected states (New York, New Jersey, Illinois, Massachusetts, California)\footnote{\url{https://www.cdc.gov/covid-data-tracker/}} than that in the least infected states (Alaska, Montana, Wyoming, Vermont). In Hawaii, people followed the stay-at-home order since March 25 and stayed longer time at home until April 20 when a significant drop was identified. The temporal curve further bounced up in late April but dropped again since May 4 when retail stores in certain parts of the state reopened. 

\section{Conclusion and Discussion}
In this work, we utilized large-scale smartphone location-derived aggregated mobility data with user-friendly design principles to develop a human mobility tracking web portal in response to the COVID-19 epidemic. We also analyzed the mobility changes across the nation regarding the statewide stay-at-home orders. High-level of adherence was found at the beginning of the orders in place although geographic variation existed. Some inevitable gathering events (e.g., in-person voting in the presidential election) were associated with mobility changes on that day and the lagged increasing COVID-19 cases two or three weeks later. It demonstrated the value of tracking human mobility patterns in the infectious disease research and in public health policy making \cite{buckee2020aggregated,oliver2020mobile}.  

It is worth noting that reduced mobility doesn't necessarily ensure the (physical) social distancing in practice according to CDC's definition\footnote{\url{https://www.cdc.gov/coronavirus/2019-ncov/prevent-getting-sick/social-distancing.html}}: "Stay at least 6 feet (2 meters) from other people". Due to the mobile phone GPS horizontal error and uncertainty \cite{gao20171}, such physical distancing patterns cannot be identified from the used aggregated mobility data; it requires other wearable sensors or trackers (e.g., Bluetooth). However, it will involve another important issue about personal data privacy and ethical concerns. It is still an ongoing challenge to find a "sweet spot" to use such data to derive effective analyses for saving lives while protecting the individual geoprivacy. We welcome user feedback from different domains for further enhancement of our web portal to inform public health decision-making practices.

\section*{Acknowledgements}
We would like to thank the Descartes Labs and the SafeGraph Inc. for providing the anonymous and aggregated human mobility and place visit data. We would also like to thank all individuals and organizations for collecting and updating the COVID-19 epidemiological data and reports. \textbf{Funding:} S.G. acknowledges the funding support provided by the National Science Foundation (Award No. BCS-2027375). Any opinions, findings, and conclusions or recommendations expressed in this material are those of the author(s) and do not necessarily reflect the views of the National Science Foundation. \textbf{Author contributions:} Research design and conceptualization: S.G., J.M.R.; Data collection and processing: S.G., Y.H.K., J.M.R.; Result analysis: S.G., J.M.R., Y.L.L; Visualization: J.M.R., J.K.; Project administration: S.G.; Writing: all authors. 

\medskip

\bibliographystyle{abbrv}
\bibliography{scibib}

\end{document}